%% file: tsocc_verif.tex
\documentclass{IEEEtran}

\usepackage{textcomp}
\usepackage{amsmath}
\usepackage{amsfonts}
\usepackage{amssymb}
\usepackage{graphicx}
\usepackage[usenames]{xcolor}
\usepackage{soul}
\usepackage{proof} 
\usepackage{listings} 
\usepackage{multirow}
\usepackage{url}
\usepackage{footmisc}

\newcommand{\trans}[1]{\stackrel{#1}{\longrightarrow}}
\newcommand{\wtrans}[1]{\stackrel{#1}{\Longrightarrow}}
\newcommand{\transstar}[1]{\mathrel{\raisebox{0ex}[1ex][0ex]{$\overset{#1}{\mathord{\longrightarrow}}$}^{*}}}

\newcommand{\CMDread}[1]{\ensuremath{\textsf{Read}(#1)}}
\newcommand{\CMDwrite}[1]{\ensuremath{\textsf{Write}(#1)}}

\newcommand{\relOrder}[1]{\mathrel{\overset{\mathsf{#1}}{\rightarrow}}}
\newcommand{\relAcyclic}[1]{\ensuremath{\mathtt{acyclic}(#1)}}
\newcommand{\relIrreflexive}[1]{\ensuremath{\mathtt{irreflexive}(#1)}}

\newcommand{\proc}[1]{\ensuremath{\textsf{proc}(#1)}}
\newcommand{\addr}[1]{\ensuremath{\textsf{addr}(#1)}}
\newcommand{\relANY}{\ensuremath{\mathbb{E} \times \mathbb{E}}}
\newcommand{\relWR}{\textsf{WR}}
\newcommand{\relWW}{\textsf{WW}}
\newcommand{\relRR}{\textsf{RR}}
\newcommand{\relRW}{\textsf{RW}}
\newcommand{\relPO}{\textsf{po} }
\newcommand{\relRF}{\textsf{rf} }
\newcommand{\relCO}{\textsf{co} }
\newcommand{\relPPO}{\textsf{ppo} }
\newcommand{\relFENCES}{\textsf{fences} }
\newcommand{\relPROP}{\textsf{prop} }
\newcommand{\relFR}{\textsf{fr }}
\newcommand{\relFRE}{\textsf{fre} }
\newcommand{\relRFE}{\textsf{rfe} }
\newcommand{\relHB}{\textsf{hb} }
\newcommand{\relCOM}{\textsf{com} }
\newcommand{\relPOLOC}{\textsf{po-loc} }

\newcommand{\murphi}{Mur$\varphi$}

\newtheorem{theorem}{Theorem}
\newtheorem{definition}{Definition}
\newtheorem{proposition}{Proposition}

\newcommand{\QA}{\mathcal{Q}_A}
\newcommand{\QC}{\mathcal{Q}_C}

\newcommand{\pair}[2]{\langle {#1}, {#2} \rangle}
\newcommand{\last}{\mathrm{last}}
\newcommand{\AbsWitness}{\mathrm{AbsWitness}}
\newcommand{\Match}{\mathrm{Match}}

\begin{document}

\title{Verification of a lazy cache coherence protocol against a weak memory model}
\author{Christopher J.\ Banks$^{1}$, Marco Elver$^{1\ast}$, Ruth Hoffmann$^{2}$, Susmit Sarkar$^{2}$, Paul Jackson$^{1}$, Vijay Nagarajan$^{1}$ \\ $^{1}$University of Edinburgh, $^{2}$University of St Andrews}

\maketitle  

\begin{abstract}
  In this paper we verify a modern lazy cache coherence protocol, TSO-CC, against the memory consistency model it was designed for, TSO. We achieve this by first showing a weak simulation relation between TSO-CC (with a fixed number of processors) and a novel finite-state operational model which exhibits the laziness of TSO-CC and satisfies TSO. 
  We then extend this by an existing parameterisation technique, allowing verification for an unlimited number of processors.
 The approach is executed entirely within a model checker, no external tool is required and very little in-depth knowledge of formal verification methods is required of the verifier. 
 
\renewcommand*{\thefootnote}{\fnsymbol{footnote}}
\footnotetext[1]{Now at Google.}
\renewcommand*{\thefootnote}{\arabic{footnote}} 
\end{abstract}

\section{Introduction}
In parallel architectures with local caches, cached values can become stale. Therefore, it is imperative that the system guarantees shared memory correctness by ensuring that it correctly implements a \emph{memory consistency model} (MCM)---the formal model that determines what value a read should return~\cite{Adve1996a}.
An integral component of enforcing an MCM is the \emph{cache coherence protocol} (CCP), which is responsible for making writes visible to other caches in an order that is consistent with the MCM. 

Traditionally, CCPs have been designed for the strictest of MCMs---\emph{sequential consistency} (SC). Previously, this has been beneficial as a way to decouple the design of a CCP from the MCM; indeed, a CCP designed for the strongest of MCMs could bolt-on to other weaker MCMs. Unfortunately, this simplicity comes at a cost. 

The strict program order requirements of SC mandates that writes are made globally visible before any subsequent memory operation from the same processor.
To guarantee this, CCPs \emph{eagerly} invalidate other non-local shared copies upon a write. In effect, such eager CCPs enforce the Single-Writer--Multiple-Reader (SWMR) invariant~\cite{sorin2011primer}---a cache line may only have either a single writer or multiple readers.  To this end, eager CCPs must maintain a vector of processors sharing a cache line, but this vector scales linearly with the number of processors~\cite{Choi2011,Ros2012}. Thus these protocols do not scale well to large-scale many-core processors.

Luckily, modern architectures tend to have more relaxed MCMs like 
Total Store Order (TSO)---used in prevalent architectures such as x86 and SPARC. Consequently, it is possible for CCP designers to take advantage of these relaxations. Indeed, there has been significant recent research on \emph{lazy coherence protocols}~\cite{Ashby2011,Choi2011,Elver2014,Ros2012}, that exploit the fact that relaxed models only require memory to be consistent at synchronisation boundaries. In these protocols, shared lines are \emph{self-invalidated} on synchronisation boundaries and therefore no longer require a (poorly scaling) sharing vector.

This poses a problem for the verification of such protocols. Traditionally, formal verification approaches for CCPs~\cite{Abts2003,Komuravelli2014} have focused on model checking 
protocol-specific  safety properties such as the SWMR invariant~\cite{sorin2011primer}. 
However, the new lazy CCPs that are designed to take advantage of weak MCMs violate SWMR by design and hence cannot be verified in the usual way.
They need to be verified in a stronger manner, for adherence to the MCM.
This is especially appropriate for the protocol we study, TSO-CC~\cite{Elver2014}, because it was designed specifically with the TSO memory model in mind.

\paragraph*{Challenges}
If the new scalable lazy CCPs are to see the light of the day, we believe they need to be formally verified against the MCM. A testing approach does not cover all corner cases and does not give the confidence that formal verification brings. Equally, the subtlety of behaviour exhibited by both lazy CCPs and weak MCMs warrants a rigorous approach. Only formal verification will suffice to allay skepticism surrounding the behaviour of lazy CCPs.
Furthermore, the verification technique should be generally applicable, should not assume the verifier to have sophisticated knowledge beyond the protocol, and it should scale to many-core processors.

\paragraph*{Our result}
In this paper, for the first time,  we  formally and exhaustively verify a modern lazy CCP
against the MCM which it is supposed to implement. Our protocol of interest is TSO-CC (Section~\ref{tso-cc}), a scalable lazy CCP which was designed to target TSO.
We establish our result for fixed cache sizes, but for any number of processors. 
Our verification focuses on safety; we do not tackle liveness. This enables our verification approach to use a slightly abstract version of CCP where, for example, access counters are not modelled explicitly.

Our approach to verification proceeds as follows. First, we propose a novel finite-state operational model TSO-LB, based on load buffers, that abstracts our lazy CCP TSO-CC. Second, we use a model checker to establish that  TSO-CC is a refinement of the TSO-LB operational model.  Initially we show refinement for a fixed number of processors;  subsequently we deploy the parameterised verification technique of Chou et al.~\cite{Chou2004} to extend our refinement result to an arbitrary number of processors. Finally, we show that the TSO-LB operational model is consistent with an axiomatic specification of TSO.

\paragraph*{Contributions} 
Our approach is inspired by Chatterjee et al.~\cite{Chatterjee2002}, who showed how CCPs can be verified against their MCMs using a model checker. Beyond this work, we make a number of specific advances. 

First, we support, for the first time, a lazy CCP through the use of a novel abstract operational model.  A lazy CCP like TSO-CC \emph{pulls} new values via self-invalidates upon a read, in contrast to conventionally eager CCPs which \emph{push} invalidates upon a write. 
The nature and timing of invalidations in eager and lazy CCPs are different. Current operational models abstract the push-based invalidates, which makes it difficult to show that lazy CCPs refine them. We therefore needed to introduce this novel operational model we call TSO-LB which abstracts 
pull-based self-invalidates.

Second, we provide a proof that our TSO-LB model satisfies an axiomatic characterisation of TSO, 
however in Chatterjee et al. the task of showing the abstract operational models are consistent with axiomatic descriptions of the MCMs is not completed (and, as far as we can tell, was never subsequently completed).  In our case, the proof is particularly important given how TSO-LB differs from conventional operational models for TSO.

Third, we employ the parameterisation technique of Chou et al.~\cite{Chou2004} to verify for an arbitrary number of processors (whereas Chatterjee et al. only verified for a fixed number of processors).  In doing so, we demonstrate that the technique is not only useful when model checking CCP properties, but also is useful when using model checking to verify refinement and show a CCP satisfies the relevant MCM.

\paragraph*{Other related work}
Another alternative approach by Manerkar et al.~\cite{Manerkar2015} uses CCICheck, which explores ordering relations between CCP and MCM; however, protocols must be described in an axiomatic style---orthogonal to typical operational descriptions of protocols---and verification is with respect to specific litmus tests---which may not capture every MCM behaviour and hence not exhaustive. It is notable that these approaches only verify for a fixed number of processors; an approach to solving this problem is found in compositional model checking approaches pioneered by McMillan~\cite{McMillan2001}. This method was further refined by Chou et al.~\cite{Chen2006,Chou2004} and made practical; however, they, once again, only deal with protocol-specific properties. Likewise, Pong and Dubois~\cite{Pong1998,Pong2000} verify compositionally, using Symbolic State Models, but again only against protocol-specific properties.
Abdulla et al.~\cite{Abdulla2016} recently propose the Dual-TSO operational model for TSO for program verification, in which they replace the store buffer in the traditional operational model with a load buffer. However, their notion of a load buffer has unbounded queues with potentially multiple values for an address, and thus does not help us with the infinite state-space problem. Our model also works very differently (but similar to CCP's like TSO-CC) by propagating multiple addresses to a load buffer atomically. So our model is not obviously a refinement of some finite restriction of the Dual-TSO model. It is also worth noting that we first defined our TSO-LB model~\cite{Elver2016thesis} concurrently with Abdulla et al.

\section{TSO-CC}\label{tso-cc}
TSO-CC~\cite{Elver2014} is a lazy CCP, designed to address the scalability issues surrounding CCPs for large numbers of cores. Lazy CCPs, like TSO-CC, take account of the fact that the relaxed memory models employed in modern multi-core processors only require memory to be consistent at synchronisation boundaries. Consequently, instead of eagerly enforcing coherence at every write, coherence is enforced lazily only at synchronisation boundaries. Thus, upon a write, data is merely written to a processor-local write-buffer, the contents of which are flushed to the shared cache upon a \emph{release}.  Upon an \emph{acquire}, shared lines in the local caches are self-invalidated---thereby ensuring that reads to shared lines fetch the up-to-date data from the shared cache. In effect, the CCP may be much simpler and \emph{does not require a sharing vector}.

However, the design of TSO-CC is specifically directed by the TSO memory model which has no explicit release or acquire instructions. It follows that, as reads have acquire semantics and writes have release semantics, a TSO compliant CCP would only need to consider each read/write an acquire/release; this, of course is not efficient because all reads and writes would need to be propagated, effectively negating the provision of local caches.

The approach in TSO-CC is that for each cache line in the shared cache, it keeps track of whether the line is exclusive, shared, or read-only. Shared lines do \emph{not require tracking of sharers} (making TSO-CC more scalable than standard directory-based protocols). Additionally, for exclusive cache lines, it only maintains a pointer to the owner.

Since it does not track sharers, writes do not eagerly invalidate shared copies in other processors. On the contrary, writes are merely propagated to the shared cache in program order (thus ensuring write-write order). To save bandwidth, instead of writing the full data block to the shared cache, it merely propagates the coherence states. Intuitively, the \emph{most recent} value of any data is maintained in the shared cache.

Reads to shared cache lines are allowed to read from the local cache, up to a predefined number of accesses (potentially causing a stale value to be read), but are forced to re-request the cache line from the shared cache after exceeding an access threshold (the implementation maintains an access counter per line). This ensures that any write (used as a release) will eventually be made visible to the matching acquire, ensuring \emph{eventual write propagation}. When a read misses in the local cache, it is forced to obtain the most recent value from the shared cache. In order to ensure the read-read order, future reads will also need to read the most recent values. To guarantee this, whenever a read misses in the local cache, it self-invalidates all shared cache lines.
Finer details of the protocol may be found in the original paper by Elver and Nagarajan~\cite{Elver2014}. It should be noted that our model implements the basic protocol, without timestamps. 

\paragraph*{Prior TSO-CC verification work}
In order to check that the protocol implementation adheres to TSO, the original 
authors of TSO-CC used the \textsf{diy}~\cite{Alglave2011} tool to generate litmus tests for TSO (according to the method detailed in Owens et al.~\cite{Owens2009}) and ran it in a full-system simulator.  An independent approach to verification was made by CCICheck~\cite{Manerkar2015}, using TSO-CC as a case study. CCICheck uses abstract axiomatic models of pipeline and memory system, and
verifies that a set of litmus tests is not violated. However, whilst a litmus test based approach provides some confidence that the protocol is correct, it is by no means an exhaustive means of verification and corner cases may be missed. In order to minimise the potential for missed corner cases in a detailed cycle-accurate full-system implementation, 
Elver and Nagarajan developed McVerSi~\cite{Elver2016}, a test generation
framework for fast memory consistency verification in simulation.
This approach, whilst it further increased confidence and testing of corner cases, is still not  exhaustive. The remainder of this paper solves this problem with an entirely exhaustive approach to verifying TSO-CC against the TSO memory model.

\section{TSO-CC satisfies TSO }\label{verify}
In this section we show that the cache coherence protocol TSO-CC does indeed satisfy the constraints of the TSO memory consistency model. This solves the problems associated with the previous verification approaches; corner cases which could be missed by insufficient testing would now be revealed by exhaustive exploration of the state space. For now, we only show that this is true for the simpler case of a fixed number of processors. We go on to show, in Section~\ref{param}, that this is true for a parameterised model of TSO-CC with any number of processors.

We took a number of discrete steps in the process of verifying the protocol. The first step was to translate the protocol into a suitable model for verification. For this purpose we chose the Mur$\varphi$ language and model checker~\cite{Dill1996}. Mur$\varphi$ is a well-established model checker and extensively used in both previous academic studies~\cite{Dill1996,Chatterjee2002,burckhardt2005verifying} and in industry~\cite{McMillan2001,Abts2003,Chou2004,OLeary2009,Choi2011}. 
We then went on to show that this model satisfied some basic properties, such as freedom from deadlock, using the model checker. Our approach to this is detailed in Section~\ref{model}.

The next step in the process was to show that the TSO-CC model satisfied the constraints of TSO. One way to achieve this was to show there exists a \emph{weak simulation relation} between TSO-CC and an operational model of TSO. A weak simulation relation exists if the observable actions (reads/writes to a memory location) in the CCP model can be matched by actions in the model of TSO. This concept is defined more formally in Section~\ref{wsim}, in which we also explain our approach to showing weak simulation using the model checker.

However, in order for our approach  
to work, we needed an operational model of TSO. Such models exist in the literature but tend to be \emph{store buffer based}~\cite{Owens2009,Sewell2010}. These models, while abstracting push-based eager CCPs well, make it difficult to show that lazy CCPs (which pull new values via self-invalidates) refine them. 
Furthermore, whereas such models require unbounded store buffers, we needed a finitely enumerable model for use with a model checking approach. Hence, in Section~\ref{tso-lb} we define TSO-LB, a \emph{load buffer based} operational model with bounded buffers that abstracts lazy CCPs. After establishing that TSO-LB exhibits only TSO behaviour, we were able to use the operational model as part of our verification strategy.

\subsection{Model checking in Mur$\varphi$}\label{model}
\begin{figure}
  \centering
  \includegraphics[width=0.4\textwidth]{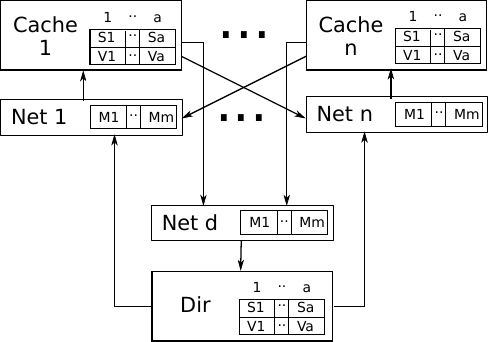}
  \caption{Model structure. The protocol model consists of a finite number of caches, $n$, each with a set of $a$ memory addresses, each with a corresponding state ($S1,\ldots,Sa$) and value ($V1,\ldots,Va$); a directory; and a network ($\text{Net} 1,\ldots,\text{Net} n, \text{and Net} d$) with a set of messages ($M1,\ldots,Mm$) waiting for each node. Arrows denote the direction of flow of messages on the network.}
  \label{fig:concrete-model}
\end{figure}

We began by defining a Mur$\varphi$ model of the TSO-CC protocol. The model implements, the basic TSO-CC protocol as described in the original paper \cite{Elver2014}, with each rule in the protocol description relating to a rule in the Mur$\varphi$ model; it has parameters for the number of processors, number of addresses, and number of values; the model was checked using three address locations and two values. The model is a faithful implementation of the protocol with the only abstraction being the abstract interpretation of the access counter---as described below. The model is constructed as a set of caches and a directory, each having a state and a set of addresses or memory locations, each with a set of possible values. There is then a set of sets of messages which act as a network; each node (cache or directory) can write or read to or from the network (Figure~\ref{fig:concrete-model}).

A set of rules, each a \emph{guard} $\implies$ \emph{action} pair, then defines the behaviour of the model. As an example, the following is a pair of sample rules taken from the full ruleset\footnote{The full ruleset can be found at \url{http://banks.ac/TSO-CC/}}:

{\footnotesize{
\begin{align*}
  c[a].\texttt{state} = \texttt{I} \implies
  & \texttt{SendGetS}(c,Dir,a); \\
  & c[a].\texttt{state} := WS
  &\text{(Read I)}\\
  \\
  c[a].\texttt{state} = \texttt{E} \implies 
  & c[a].\texttt{val} := v \\
  & c[a].\texttt{state} := M; 
  & \text{(Write E)}
\end{align*}}}

where $c$ is a cache, $a$ is an address (memory location), $v$ is a value, and $c[a].\texttt{state}$ (or $c[a].\texttt{val}$) is the state (or value) for the given cache and address. The first rule (Read \texttt{I}) is the Read rule for the \texttt{I}nvalid cache state and the second is the Write rule for the \texttt{E}xclusive cache state. When a cache is in state \texttt{I} and does a Read, it sends a \texttt{GetS} message to the directory and switches to state \texttt{WS}. When a cache is in state \texttt{E} it may do a Write, store the written value, and switch to state \texttt{M} (\texttt{M}odified). The function \texttt{SendGetS} handles the passing of a \texttt{GetS} message to the network.

We then define a rule that handles the receipt of messages from the network at each node (cache or directory). Within this rule are some functions which handle actions performed when a message is received; the following is an extract from the \texttt{DirectoryReceive} function for handling the messages in the previous example:

{\footnotesize{
$\texttt{DirectoryReceive}(msg,a) =$
\begin{align*}
  \textbf{ if } & Dir[a].\texttt{state} = \texttt{I} \land msg.\texttt{type} = \texttt{GetS} \\
  \textbf{ then } 
                & \texttt{SendDataS}(msg.\texttt{src},a,\ldots); \\
                & \texttt{ReplaceOwner}(msg.\texttt{src},a); \\
                & Dir[a].\texttt{state} := \texttt{WE1} \\
  \textbf{ else } & \ldots
\end{align*}}}

There is a function \texttt{CacheReceive} which has similar conditions for receiving messages at a cache.

Another pair of rules which are of interest are the Read rules for the Shared cache state. Part of the lazy invalidation scheme for TSO-CC is that a cache must self-invalidate after a certain number of reads, once an access count reaches a certain limit. In our model we abstract away from the access counter and just have two rules corresponding to a Read in the Shared state: one where the access count is within its limit and another for when the limit has bean reached. There is thus a non-deterministic choice between the two options:

{\footnotesize{
\begin{align*}
  c[a].\texttt{state} = \texttt{S} \implies
  & \texttt{SendGetS}(c,Dir,a); \\
  & c[a].\texttt{state} := WS \\
  &\hspace{7em}\text{(Read S(MAX))}\\
  \\
  c[a].\texttt{state} = \texttt{S} \implies
  & \text{//do nothing} \\
  &\hspace{7em}\text{(Read S($<$MAX))}
\end{align*}}}

In the first rule the access count has been reached, so the cache self-invalidates and asks the directory for the fresh value; in the second rule the access count has not been reached, so the cache is free to read its own value and nothing else needs to be done. The model checker accounts for the non-deterministic choice between these rules.

Once the full ruleset in the model checker is defined, the rules are then exhaustively applied using an appropriate strategy (e.g.\ breadth first, depth first) until every possible state of the model has been enumerated; during this process of state enumeration the model checker checks that it can always proceed to another state (deadlock freedom) and that any defined invariants hold for each state. For efficiency, Mur$\varphi$ also reduces the set of states which need to be enumerated by using various techniques, such as symmetry reduction~\cite{NorrisIP1996}.

The next problem was to decide what property to check the model against. The derived properties which usually hold for CCPs, like SWMR do not hold for TSO-CC, by design, so a new strategy has to be applied. Our verification strategy is to establish that TSO-CC satisfies TSO by: (a) devising TSO-LB, a finite operational model for abstracting TSO-CC (b) proving that TSO-LB shows only TSO behaviour and (c) showing that TSO-CC is a refinement of TSO-LB within a model checker.

\subsection{TSO-LB operational model}\label{tso-lb}
This section introduces the abstract TSO load-buffering model (TSO-LB). For our approach, existing operational models of TSO~\cite{Owens2009,Sewell2010} are not ideal for two reasons. The first being that they require unbounded buffers, making algorithmic verification difficult. 
Second, a refinement between a lazy CCP and an existing store-buffering model would be difficult,  
as a lazy CCP effectively follows a load-buffering rather than a store-buffering approach: loads, viz. reads, hitting on a locally ``buffered'' (potentially) stale value, until the current value is pulled in (i.e. propagates) from global memory via a self-invalidate. 
The load-buffering based operational model formalised below abstracts a lazy CCP better and hence simplifies verification.

\begin{definition}[Labelled Transiton System]
A \emph{labelled transition system} (LTS) is a tuple $(\mathcal{L},\mathcal{Q},\mathcal{I},\mathcal{T})$ where, $\mathcal{L}$ is a set of \emph{labels}, $\mathcal{Q}$ is a set of \emph{states}, $\mathcal{I}\subseteq \mathcal{Q}$ is a set of initial states and $\mathcal{T} \subseteq \mathcal{Q} \times \mathcal{L} \times \mathcal{Q}$ is the \emph{transition relation}.
\end{definition}
If $(q,l,q')\in\mathcal{T}$ then we say there is a transition labelled $l\in\mathcal{L}$ from state $q\in\mathcal{Q}$ to state $q'\in\mathcal{Q}$ and we may abbreviate this as $q\trans{l}q'$.

\begin{definition}[TSO-LB]\label{def:tsolb}
We define an LTS for TSO-LB as follows.  The transition relation is given by the rules: 
{\small{
\[
  \infer[\textsc{\footnotesize Read}] 
  {q \xrightarrow{\CMDread{p, a, v}} q}
  {\mathtt{local}_{q}(p)(a) = v} 
\]
\[
  \infer[\textsc{\footnotesize Write}] 
  {q \xrightarrow{\CMDwrite{p, a, v}}  
      \langle \mathtt{local}_{q}[(p)(a) \mapsto v],\, \mathtt{global}_{q}[(a) \mapsto v]\rangle } 
  {}
\]
\[
  \infer[\textsc{\footnotesize Propagate}] 
  {q \xrightarrow{\tau} 
      \langle \mathtt{local}_{q}[(p) \mapsto \mathtt{global}_{q}],\, \mathtt{global}_{q} \rangle} 
  {}
\]}}
\noindent{}where $P$ is a finite set of processors, with $p \in P$; $A$ is a finite set of addresses (memory locations), with $a \in A$; $V$ is a finite set of data values, with  $v \in V$; $\mathtt{local}_q : P \rightarrow A \rightarrow V$ is a function where $\mathtt{local_q}(p)(a)$ is the value at address $a$ in the local buffer of $p$ in state $q$ and $\mathtt{global}_q : A \rightarrow V$ is a function where $\mathtt{global}_q(a)$ is the value at address $a$ in the global buffer in state $q$. 

The set of states $\mathcal{Q}$ consists of all pairs $\langle \mathtt{local}_{q}, \mathtt{global}_{q} \rangle$ and the set of labels $\mathcal{L}=\{\tau,\CMDread{p,a,v},\CMDwrite{p,a,v}\}$ where $p\in P$, $a\in A$, $v\in V$ and $\tau$ is the silent action. We define the set of initial states $\mathcal{I}$ to be $\mathcal{I} \triangleq \{q\ :\ \forall p \in P.\ \forall a \in A.\ \mathtt{local}_{q}(p)(a) = \mathtt{global}_{q}(a)\}$.

\end{definition}

\subsection{TSO-LB satisfies TSO}
We prove (Appendix~\ref{apx:proof}) that the operational TSO-LB model defined above in Definition~\ref{def:tsolb} exhibits only TSO behaviour. Since TSO-LB is defined as a LTS, behaviour of TSO-LB is defined with respect to an arbitrary trace of this LTS. We show (Theorem~\ref{thm:tsolb_is_tso}), by means of an interpretation of logical and physical time over these traces, that the behaviour satisfies the \texttt{herd} axiomatic characterisation of TSO~\cite{Alglave2014a}. We also show via a counterexample (Appendix~\ref{apx:counter}) that TSO-LB does not permit all allowable behaviours of TSO, i.e. TSO-LB is in fact stricter than TSO.

\begin{theorem}[TSO-LB satisfies TSO]\label{thm:tsolb_is_tso}
The read and write events of traces of the TSO-LB LTS satisfy the TSO axiomatic memory consistency model (as formalised in Alglave et al. \cite{Alglave2014a}, and recalled in Appendix~\ref{apx:framework}).
\end{theorem}

Our proof strategy starts with defining a trace $P$ of TSO-LB (Definition~\ref{def:trace}). The trace order might be seen as the physical-time representation of events, which contains writes, reads, and propagates. We will then construct a \emph{strict linear order} $L$ from $P$ which contains the same writes and reads (with the same values). We will then show how to instantiate the required ordering relations from the \texttt{herd} framework of Alglave et al.~\cite{Alglave2014a} (see Table~\ref{table:mcm_relations} in Appendix~\ref{apx:framework}) from $L$, and show all those orders
are contained in $L$. This will then allow us to show that the \texttt{herd} axiomatic  constraints of TSO 
hold over the write and read events. Note that we assume a simplified TSO model excluding fences, as TSO-LB does not model fences by definition.

\begin{definition}[Trace]\label{def:trace}
  A \emph{trace} of an LTS is a sequence (finite or infinite) of labels that results from a path of transitions starting at the initial state. 
  Let us call this trace order $P$.
\end{definition}

\begin{definition}[Logical-time $L$]\label{logtime}
  We define $L$ to be an order on the read and write events in the trace $P$. All writes $\CMDwrite{p,a,v}$ appear in $L$ in the same order as in the physical-time trace $P$. A read $\CMDread{p,a,v}$ is pulled backwards in the trace to just after the event 
  in $P$ which made the processor $p$ get the value $v$ for $a$. Such an event is either a write from the same processor $p$, or a propagate to the processor $p$ (if $\CMDread{p,a,v}$ reads from a write on another processor). 
\end{definition}

Note that several reads from the same processor can be pulled back to the same point in this scheme, if the same address is read by multiple reads, or if the propagated values for different addresses for the same propagate event are read from by different reads. In such a case, we order these multiple reads in $L$ (which have to be from the same processor) according to program order.

\begin{definition}[\relCO in TSO-LB]\label{cotso}
  The order \relCO is defined in TSO-LB as $\CMDwrite{p,a,v} \relOrder{\relCO} \CMDwrite{q,b,w}$ if and only if $\CMDwrite{p,a,v}$ occurs before $\CMDwrite{q,b,w}$ in the physical-time trace $P$, and the addresses $a$ and $b$ are the same. 
\end{definition}
Note that $p$ and $q$ may be the same or different processors, and $v$ and $w$ the same or different values.

\begin{definition}[\relRF in TSO-LB]\label{rftso}
  The order \relRF is defined as $\CMDwrite{p,a,v} \relOrder{\relRF} \CMDread{q,a,v}$ where $p$ and $q$ may be the same or different processors, and the read gets its value from the write.
\end{definition}
  
We can now show that \relCO, \relRF, and all the derived relations of the \texttt{herd} TSO formalisation are sub-orders of $L$.
Then all axioms state the acyclicity and irreflexivities of various order relations, which are satisfied by any sub-orders of a strict linear order $L$. For the complete proof refer to Appendix~\ref{apx:proof}. 

\subsection{Weak simulation by model checking}\label{wsim}
Our core goal here is to check that a value read from a memory location by a processor at any point in time adheres to the TSO-LB specification, if all memory accesses are governed by the TSO-CC protocol.

We model both the TSO-LB specification and the TSO-CC protocol as labelled transition systems.  In both cases, the labels are either \emph{observable actions} concerning reads and writes or they are \emph{silent actions}.  For convenience below, we use the single label $\tau$ for all silent actions, though in our implementation it is useful to consider each system having a number of silent actions.

Our formal notion of correctness is that every observable trace of the TSO-CC protocol LTS is also an observable trace 
of the TSO-LB specification LTS.
An \emph{observable trace} is a trace with all the silent actions removed.
We establish this inclusion property of observable traces by exhibiting a weak simulation relation between the TSO-CC LTS and the TSO-LB LTS such that the pair of initial states of the two LTSs is included in the relation.

A \emph{weak simulation relation} shows step-by-step that for every observable action in TSO-CC there is a corresponding observable action in TSO-LB; it makes no attempt to match the silent actions in the two LTSs.
This notion of weak simulation may be defined more formally as follows (following Milner~\cite{Milner1999}).

\begin{definition}[Weak transition] 
  Let $\mathcal{A}=(\mathcal{L},\mathcal{Q},\mathcal{T})$ be an LTS.
  A \emph{weak transition} $q\wtrans{l}q'$ is defined as $q\transstar{\tau}x\trans{l}y\transstar{\tau}q'$ for some $x$,$y$, where $\transstar{\tau}$ is the reflexive transitive closure of $\trans{\tau}$ 
  and $q,q',x,y\in\mathcal{Q}$, $l\in\mathcal{L}$ and $l\neq\tau$.
\end{definition}
Later we use the notation $q \Longrightarrow q'$ for $q\transstar{\tau}q'$ or, if we allow multiple silent-action labels, to say that $q'$ can be reached from $q$ by zero or more transitions labelled by silent actions.

\begin{definition}[Weak simulation]\label{wsimdef}
  Let $\mathcal{C}=(\mathcal{L},\QC,\mathcal{T}_{C})$ and $\mathcal{A}=(\mathcal{L}, \QA, \mathcal{T}_{A})$ be two LTSs with the same label set.
  Let $l \in \mathcal{L}$ be an observable action.
  A \emph{weak simulation} $\mathcal{W} \subseteq \QC \times \QA$ is a binary relation such that if $(p,q)\in\mathcal{W}$, written $p\mathcal{W}q$, then
  \begin{enumerate}
  \item if $p\trans{l}p'$ then there exists $q' \in \QA$ such that $q \wtrans{l}q'$ and $p'\mathcal{W}q'$, and
  \item if $p\trans{\tau}p'$ then there exists $q' \in \QA$ such that $q \Longrightarrow q'$ and $p'\mathcal{W}q'$.
  \end{enumerate}
\end{definition}
In our setting, $\QC$ are the states of the CCP and $\QA$ are the states of the MCM.

To prove that there exists a weak simulation relation using a model checker, we construct an unlabelled transition system $\mathcal{M} = (\mathcal{Q},\mathcal{T})$ from the two LTSs with $\mathcal{Q}=\QC \times \QA$ and a specially crafted transition relation $\mathcal{T}$. 
If a certain property holds for every reachable state of $\mathcal{M}$, then the set of reachable states is a weak simulation relation between $\mathcal C$ and $\mathcal A$.
As the initial state of $\mathcal M$ is a pair of the initial states of $\mathcal C$ and $\mathcal A$, we have that the initial state pair are related by the weak simulation, and hence every observable trace of $\mathcal C$ is also an observable trace of $\mathcal A$.
We can describe the transition relation and checked property as follows. The transition relation $\pair{p}{q} \longrightarrow \pair{p'}{q'}$ is defined as $\exists l \in L.\,  p \trans{l} p' \land q' = \last(\AbsWitness(p,q,l))$ where $\AbsWitness(p,q,l)$ computes an alternating sequence of abstract states and labels $\langle q_0,l_0,q_1,l_1,\ldots, q_n\rangle$ for some $n \geq 0$, $q = q_0$ and the $\last()$ function picks out the last state $q_n$ of such a sequence.

The checked property $\Match(\pair{p}{q})$ is defined as ${\forall l \in L, p' \in \QC.\, p\trans{l}p'} \Rightarrow \AbsWitness(p,q,l)$ is a witness for:
  \begin{enumerate}
  \item $q\wtrans{l} \last(\AbsWitness(p,q,l))$ if $l$ is observable. and
  \item $q\wtrans{} \last(\AbsWitness(p,q,l))$ if $l=\tau$.
  \end{enumerate}
 
Here, an alternating sequence of abstract states and labels $\langle q_0,l_0,q_1,l_1,\ldots, q_n\rangle$ \emph{is a witness for} $q_0\wtrans{}q_n$ if all the $l_i$ are silent and $q_i\trans{l_i}q_{i+1}$ for all $i \in \{0,\ldots,n-1\}$, and \emph{is  a witness for} $q_0\wtrans{l}q_n$ if there exists a unique $l_{i}=l$ in the sequence such that $\forall j \neq i$ $l_{j}$ is silent and  $q_i\trans{l_i}q_{i+1}$ for all $i \in \{0,\ldots,n-1\}$.

Witnesses for weak transition instances enable the straightforward checking of the truth of instances.

A conceptual sketch of the witness function $\AbsWitness$ we use 
is as follows:
\begin{itemize}
\item If TSO-CC does a write action, then TSO-LB is made to take a single corresponding write action step.
\item For
  silent transitions of TSO-CC, the witness is a single state -- i.e. TSO-LB takes no steps. 
\item If TSO-CC does a read action, then in TSO-LB we either do a read action or do a propagate action followed by a read action.  We settle for the single read step if it is allowed by the TSO-LB.  If not, we go for the 2 step witness.  As propagate is the only silent action in TSO-LB and it is idempotent, there are no other options to consider.
\end{itemize}
In general the abstract LTS might permit several silent transitions and the $\AbsWitness$ function has to embody some strategy for testing possible silent actions; however, it is worthy of note that the trivial strategy, as described here, is generally applicable to checking \emph{any} CCP against TSO-LB.

\subsection{Weak simulation in Mur$\varphi$}
To realise the above in Mur$\varphi$ we started with the Mur$\varphi$ TSO-CC model introduced above in Section~\ref{model} and augmented the state with components for the TSO-LB specification.
At the rules in the TSO-CC model where the observable actions (reads/writes) are performed, we also step forward the TSO-LB model with the same observable actions, as explained above.  

Coding the $\Match$ predicate is much simpler than the conceptual presentation above suggests. 
For the step forward of the TSO-LB system on write actions, the step is guaranteed by construction to satisfy the TSO-LB labelled transition relation, there is nothing to check.  Only for the read action do we need to check that the value read in the TSO-LB specification actually matches that from the TSO-CC system. We simply use an invariant in Mur$\varphi$ to check the read value at each read step.

We now give some further concrete details of how we coded the transition system model and $\Match$ check in Mur$\varphi$.
The implementation of TSO-LB involves a pair of arrays to represent the global and local buffers for each cache and address, \murphi{} procedures \texttt{TSOStore} and \texttt{TSOUpdate}, and the \murphi{} function \texttt{TSOVerify}. These functions compute the next TSO-LB state for the Write, Propagate, and Read TSO-LB rules respectively. In addition \texttt{TSOVerify} returns a Boolean value indicating whether TSO-LB can indeed make one or two steps forward that result in a correct read. Calling \texttt{TSOVerify} returns true if the expected value is in the local buffer, or it tries a \texttt{TSOUpdate} and returns true if the expected value is now in the local buffer, else it returns false. A Mur$\varphi$ invariant ensures that \texttt{TSOVerify} always returns true.

The rules of TSO-CC incorporate these TSO-LB procedures and function. Taking our previous example rules, we amend them as follows:

{\footnotesize{
\begin{align*}
  c[a].\texttt{state} = \texttt{E} \implies 
  & c[a].\texttt{val} := v \\
  & c[a].\texttt{state} := M; \\
  & \fbox{\texttt{TSOStore}$(c,a,v)$}\\
  & \hspace{8em}\text{(Write E)} 
  \\\\
  c[a].\texttt{state} = \texttt{S} \implies
  & \text{//do nothing}; \\
  & \fbox{\texttt{Assert}(\texttt{TSOVerify}($c,a,c[a]$.\texttt{val}))}\\
  & \hspace{8em}\text{(Read S($<$MAX))}\\
\end{align*}}}

and likewise wherever a Read or Write action occurs in the CCP model. In this way our model shows that the values at the CCP level are consistent with the values at the MCM level. 

Thus, for a fixed number of processors, we show that the simulation relation between TSO-CC and TSO-LB holds. The next problem was to show that the simulation relation holds for any number of processors. The next section shows how we solved this problem.

\section{TSO-CC with $n$ processors satisfies TSO}\label{param}
After showing that TSO-CC indeed satisfies TSO for a finite number of processors, we now show that this is also the case irrespective of the number of processors. In this section we present a parameterised model, parameterised in the number of processors, showing the same weak simulation relation between TSO-CC and TSO still applies with $n$ processors.

\begin{figure}
  \centering
  \includegraphics[width=0.4\textwidth]{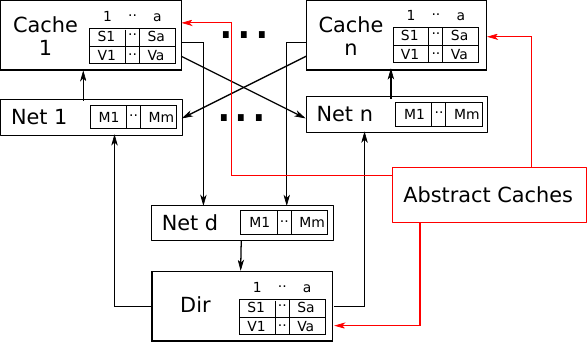}
  \caption{Parameterised model structure with abstract caches. This consists of a finite number of caches, $n$, each with a set of $a$ memory addresses, each with a corresponding state ($S1,\ldots,Sa$) and value ($V1,\ldots,Va$); a directory; and a network ($\text{Net} 1,\ldots,\text{Net} n, \text{and Net} d$) with a set of messages ($M1,\ldots,Mm$) waiting for each node. Arrows denote the direction of flow of messages on the network. There is then an abstract cache which can arbitrarily send messages to the concrete nodes.}
  \label{fig:abstract-model}
\end{figure}

In order to define a parameterised model we follow the method of Chou et al.~\cite{Chou2004}, who in turn refined the ideas of McMillan~\cite{McMillan1999}; the method is proven mathematically correct by Krsti\'{c}~\cite{Krstic2005}. The essence of the method is that one takes the original concrete model, but adds a new \emph{abstract cache}. The abstract cache represents any number of caches connected to the concrete model (Figure~\ref{fig:abstract-model}). Initially the abstract cache can send any possible message to the concrete caches. This over-approximated set of messages is then reduced to only the set of legal messages by a process of counterexample guided abstraction refinement~\cite{Clarke2003}.

The initial over-approximated set of messages coming from the abstract cache will generate a counterexample when a spurious message is sent. One can then introduce a restriction to the abstract cache which disallows the spurious message. However it is then necessary to show that the restriction is valid and does not lead to an under-approximation of legal messages. In order to achieve this, one can write a \emph{non-interference lemma} which shows the restricted message cannot occur in the concrete model. The key to the method is that the 
process is manual, but simply mechanical: the restriction is guided by the counterexample, the lemma is guided by the restriction, then the model checker checks both simultaneously and automatically. The apparent circular reasoning in proving the lemma on the amended system is dealt with by example by Chou et al.~\cite{Chou2004} and proved correct by Krsti\'{c}~\cite{Krstic2005}.

\subsection{Parameterised model in Mur$\varphi$}
The implementation of the parameterised model begins with the definition of a set of rules which generate all the possible messages which could be received at each node from the abstract caches. For example, one of the rules which handles the receipt of a \texttt{DataX} message at a cache is:

{\footnotesize{
\begin{align*}
  c[a].\texttt{state}=\texttt{WX} \\
  \implies & \texttt{SendAck}(c,Dir,a); \\
           & c[a].\texttt{state} := \texttt{M}; \\
           & \texttt{InvalidateAllOtherLines}(c,a) \\
           & \hspace{7em}\text{(Cache Recv DataX Abs)}
\end{align*}}}

and similar rules are defined for each combination of node, state, and message received.

To extend the MCM to the parameterised model, we must consider what happens when our observable actions are performed by the abstract part of the parameterised model. As we do not track the state of abstract processors a Read action is not explicitly defined in the abstract part of the model, however a Write action by an abstract processor would have an effect (eventually) on the state of the concrete caches. We do not keep track of values at the abstract caches, so local buffers for abstract caches are not needed in the memory model. However, a Write at an abstract cache will go to the global buffer, because it may at some point be read by a concrete cache.

Thus, we implement a new function, \texttt{TSOStoreAbs}, which writes only to the global buffer. Now, in our abstract cache rules, we add a call to \texttt{TSOStoreAbs} wherever we see a Write action. For example, when the directory receives a Data message from an abstract cache a Write has occurred and we record this in the global buffer:

{\footnotesize{
\begin{align*}
  c[a].\texttt{state}=\texttt{WS} 
  \implies & dir[a].\texttt{val} := v; \\
           & \texttt{TSOStoreAbs}(a,v) \\
           & dir[a].\texttt{state} := \texttt{S} \\
           & \hspace{6em}\text{(Dir Recv Data Abs)}
\end{align*}}}

Once these rules are defined the model checker will generate all possible messages coming from the abstract cache. At this stage we have an \emph{over-approximation} of the system.  Of course, some of these messages will not be valid in the current state. When this occurs a counterexample will be generated by the model checker. The modeller must then inspect the counterexample and work out why the message was spurious. It is then possible to add a restriction to the rule that generated it such that the spurious message is eliminated. 

For example, in the above rule (Cache Recv DataX Abs), we allow the cache to receive a \texttt{DataX} even when it is not the owner of the cache line. This produces a counterexample, because to receive a \texttt{DataX} from another cache (here an abstract cache) the other cache must have received a \texttt{FwdX} message first telling it to forward data to the new owner. Therefore the receiving cache must be the owner. To eradicate the counterexample we must add a restriction to check the receiving cache is the owner:

{\footnotesize{
\begin{align*}
  c[a].\texttt{state}=\texttt{WX } & \fbox{$\land$ \texttt{ IsOwner}$(c,a)$} \\
  \implies & \texttt{SendAck}(c,Dir,a); \\
           & c[a].\texttt{state} := \texttt{M}; \\
           & \texttt{InvalidateAllOtherLines}(c,a) \\
           & \hspace{6em}\text{(Cache Recv DataX Abs)}
\end{align*}}}

However, we must now show that the restriction is not too strict, i.e.\ we have not inadvertently caused the system to be \emph{under-approximated} and, in essence, we are not changing the protocol. To do this, we introduce a \emph{non-interference lemma}; this is a lemma which states the restriction as an invariant in the context of the concrete model, thus ensuring that the spurious messages eliminated are indeed not possible in the fully concrete model. For example, the lemma for the above restriction is:

{\footnotesize{
\begin{align*}
  \forall n \forall a \forall i. \hspace{0.5em} net[n][a][i].\texttt{msgType}=\texttt{DataX} 
  \implies \texttt{IsOwner}(n,a) 
\end{align*}}}

where $n$ is a node, $a$ is an address, and $i$ is a position in the message buffer. This is implemented in the model checker as an invariant\footnote{Details of the restrictions and non-interference lemmas can be found in the model source at \url{http://banks.ac/TSO-CC/}} and if it does not fail then we know that we have not over-constrained the abstract cache.

Now, the model checker may catch a new counterexample. If this is the case then we repeat the process until all counterexamples are eliminated. Once all counterexamples are eliminated, we are done.

\section{Results}
In summary, the result of applying the method described in this paper to the TSO-CC protocol was that we showed that the protocol does, in fact, conform to the TSO memory model with any number of processors. 
Execution times for checking the full model are in the order of 14--15 hours on a single core of an Intel Xeon 1.8GHz machine with 64GB of RAM. The process of manually refining the model for parameterisation required 30 passes around the refinement loop, generating 30 non-interference lemmas. The time needed to define each lemma varied, depending on the complexity of the counterexample---at this stage, detailed knowledge of the protocol was a boon. Of note, however, is that for each pass around the refinement loop does not require 14 hours of model checking; generally, the model checker needed only to run for a few minutes to find the next counterexample---this time gradually increased as more counterexamples were eliminated.

\section{Conclusion}
We have shown that it is possible to verify a modern, lazy CCP against its counterpart TSO MCM. 
Our main contributions have been three fold: 

The extension of a previous method~\cite{Chatterjee2002} in order to formally verify a lazy CCP against the TSO weak MCM that it implements.  The key novelty that enables this extension is the introduction of the new abstract operational model TSO-LB.

A proof that our TSO-LB model satisfies a well-regarded axiomatic characterisation of TSO.

The extension of our verification result to an arbitrary number of processors, using the parameterised verification technique of Chou et al~\cite{Chou2004}. In establishing this result, we show this technique can be used to show a CCP refines an abstract operational model, not just for verifying protocol-specific properties such as SWMR.

We believe it would be straightforward to use our approach to verify other lazy CCPs that implement TSO.
One direction of future work is to improve the degree of automation in the method. Whilst the process of parameterisation of the model is simple, requiring more knowledge of the protocol than of formal methods, of note is the time and effort required to write restrictions, write lemmas, model check, and repeat. It is our belief that more of this process may be automated, as was the goal of both Chou et al.~\cite{Chou2004} and Krsti\'{c}~\cite{Krstic2005}. Some research on this topic already exists in the literature, for example O'Leary et al.~\cite{OLeary2009} and Bingham et al.~\cite{Bingham2008}. Another direction for future work is to check how we might use results similar to those of Henzinger et al.~\cite{Henzinger99} to justify the verification for arbitrary numbers of addresses and data values.
\input{tsocc_verif.bbl}

\appendix

\section{Summary of Axiomatic Framework}\label{apx:framework}
This section summarises the framework proposed by Alglave et al.~\cite{Alglave2014a} to specify the axiomatic semantics of MCMs. We have chosen to use this framework, as it is succinct while eliminating ambiguity (especially for the purpose of implementing decision procedures) and is flexible enough to describe a variety of consistency models while relying on a library of common reusable definitions (hence a framework).

\begin{definition}[Events]
An \emph{event} captures the thread (\proc{e}), address (\addr{e}) and action. Events are unique, referred to with a lower case letter (e.g. $e$). For all events in a program, \relWR, \relWW, \relRR, \relRW are the relations capturing all write-read, write-write, read-read and read-write pairs respectively.
\end{definition}

\begin{definition}[Candidate executions]\label{definition:mcms:candidate_exec}
A \emph{candidate execution} is a tuple $(\mathbb{E}, \relPO, \relRF, \relCO)$, with the set of events $\mathbb{E}$ in the program, and the relations program-order \relPO (via control-flow semantics), reads-from \relRF and coherence (or write-serialisation) order \relCO (via data-flow semantics); see Table~\ref{table:mcm_relations} for details.
\end{definition}

\subsection{Architecture Definition}
An \emph{architecture} defines the architecture specific details of a particular memory consistency model, and is used by the constraint specification to decide if a particular execution is valid or not. An architecture is defined by the tuple of relations $(\relPPO, \relFENCES, \relPROP)$, preserved program-order \relPPO, \relFENCES and propagation \relPROP{}; see Table~\ref{table:mcm_relations} for details.

\begin{definition}[Total Store Order]\label{definition:mcms:arch_tso}
  Only the write to read ordering is relaxed. Reads to the same address as a preceding writes $w$ by the same thread must observe either $w$ or a write by another thread that happened after $w$ in \relCO; this, however, has no effect on the required visibility by other threads, which implies that \relPROP only includes \relRFE (and not \relRF). The relation \textsf{mfence} captures write-read pairs separated by a fence instruction.
  \[
    \begin{array}{l l}
      \relPPO &\triangleq \relPO \setminus \relWR\\
      \relFENCES &\triangleq \textsf{mfence}\\
      \relPROP &\triangleq \relPPO \cup \relFENCES \cup \relRFE \cup \relFR
    \end{array}
  \]
\end{definition}

\begin{table}
  \centering
  \caption{Definition of relations used to specify memory consistency
  models in the axiomatic framework.}
  \label{table:mcm_relations}
  \scriptsize
  \begin{tabular}{| c | p{0.16\linewidth} | l | l | p{0.28\linewidth} |}
  \hline
  \emph{Relation} & \emph{Name} & \emph{Source} & \emph{Subset of} & \emph{Definition}
  \\\hline\hline
  \relPO & program-order & execution & \relANY & instruction order lifted to events \\\hline
  \relRF & read-from & execution & \relWR & links a write $w$ to a read $r$ taking its value from $w$ \\\hline
  \relCO & coherence & execution & \relWW & total order over writes to the same memory location\\\hline\hline
  \relPPO & preserved program order & architecture & \relPO & program order maintained by the architecture \\\hline
  \relFENCES & fences & architecture & \relPO & events ordered by fences \\\hline
  \relPROP & propagation & architecture & \relWW & order in which writes propagate\\\hline\hline
  \relPOLOC & program order subset of same address events & derived & \relPO & \(\relPOLOC{} \triangleq\)\newline
    \(\{(x, y)\ |\ x \relOrder{\relPO} y \land \addr{x} = \addr{y}\}\)
  \\\hline
  \relCOM & communications or conflict orders & derived & \relANY & $\relCOM \triangleq \relCO \cup \relRF \cup \relFR$ \\\hline
  \relRFE & read-from external & derived & \relRF & $\relRFE \triangleq$\newline 
$\{(w, r)\ |\ w \relOrder{\relRF} r \land \proc{w} \neq \proc{r}\}$ \\\hline
  \relFR & from-read & derived & \relRW & links reads to writes based on observed \relRF{} and \relCO: $\relFR \triangleq (\relRF^{-1};\relCO) \triangleq \{(r, w)\ |\ \exists w'. w' \relOrder{\relRF} r \land w' \relOrder{\relCO} w\}$ \\\hline
  \relFRE & from-read external & derived & \relFR & $\relFRE
    \triangleq$\newline
    $\{(r, w)\ |\ r \relOrder{\relFR} w \land \proc{r} \neq \proc{w}\}$ \\\hline
  \relHB & happens before & derived & \relANY & $\relHB \triangleq \relPPO \cup \relFENCES \cup \relRFE $\\\hline
  \end{tabular}
\end{table}

\subsection{Constraint Specifications}
\label{section:mcms:constraints}

Given a candidate execution and an architecture specification, the constraints (or axioms) decide if the execution is valid under the complete model.

\begin{definition}[\sc SC PER LOCATION]\label{def:sc_per_location}
\textsc{SC PER LOCATION} ensures that communications/conflict orders \relCOM cannot contradict program-order per memory location \relPOLOC. Formally \textsc{SC PER LOCATION} is satisfied if
\[
    \relAcyclic{\relPOLOC \cup \relCOM}.
\] 
\end{definition}

\begin{definition}[\sc NO THIN AIR]\label{def:no_thin_air} 
This constraint ensures that the happens-before order \relHB, which captures preserved program-order \relPPO, fenced instructions \relFENCES, but also reads from other threads \relRFE is not contradictory. Formally \textsc{NO THIN AIR} is satisfied if
\[
    \relAcyclic{\relHB}.
\]
Effectively, this prevents reads from observing values ``out of thin air'', i.e.  before they appear to have been written by some other thread.
\end{definition}

\begin{definition}[\sc OBSERVATION]\label{def:observation}
\textsc{OBSERVATION} constrains the values a read may observe. Assume a write $w_a$ to address $a$, a write $w_b$ to address $b$, which are ordered in \relPROP{} ($w_a \relOrder{\relPROP} w_b$), and a read $r_b$ reading from $w_b$ ($w_b \relOrder{\relRF} r_b$), then any read $r_a$ that happens after $r_b$ ($r_b \relOrder{\relHB} r_a$) cannot read from a write before $w_a$. \textsc{OBSERVATION} is satisfied if
\[
    \relIrreflexive{\relFRE;\relPROP;\relHB^*}.
\]
\end{definition}

\begin{definition}[\sc PROPAGATION]\label{def:propagation}
This constraint imposes restrictions on the order in which writes are propagated to other threads, i.e. ensuring that the propagation order \relPROP does not contradict coherence (write-serialisation) order \relCO. This constraint and depending on the \relPROP order defines if the consistency model is \emph{write/multi-copy atomic}, i.e. if all threads observe writes in the same order or not.
\textsc{PROPAGATION} is satisfied if
\[
    \relAcyclic{\relCO \cup \relPROP}
\].
\end{definition}

\subsection{Proof that TSO-LB satisfies TSO}\label{apx:proof}

\begin{proposition}
  The \relCO order defined in Definition~\ref{cotso} is a sub-order of the logical time order $L$ from Definition~\ref{logtime}.
\end{proposition}
\begin{IEEEproof}
  Since $L$ has writes in the same order as $P$, if $\CMDwrite{p,a,v}$ is before $\CMDwrite{q,b,w}$ in $P$, it is before in $L$ as well.
\end{IEEEproof}

\begin{proposition}
  The \relRF order defined in Definition~\ref{rftso} is a sub-order of the logical time order $L$ from Definition~\ref{logtime}.
\end{proposition}
\begin{IEEEproof}
  There are two cases, where the read and the write are from the same or different processors.
  
  If the read and the write are from the same processor, the read has been pulled to be just after the write in $L$, so the write is before the read.

  If the read and the write are from different processor, in $P$ there must have been a propagate to the processor of the read before the read. This propagate must necessarily be after the write in $P$ (to get its value). In constructing $L$, we have pulled the read to just after that propagate, so the write is still before the read.
\end{IEEEproof}

\begin{proposition}\label{frtso}
  By defining $\relFR \triangleq (\relRF^{-1};\relCO)$ as usual, \relFR is a sub-order of the logical time order $L$ from Definition~\ref{logtime}.
\end{proposition}
\begin{IEEEproof}
  Consider a $\CMDread{p,a,v} \relOrder{\relFR} \CMDwrite{q,a,w}$. There are two cases, where the read and the write are from the same or different processors.

  If the read and the write are from the same processor, the read must come before the write in program order, and therefore before in $P$. Since reads are only pulled back in constructing $L$, the read is still before the write in $L$.

  If the read and the write are from different processors, the write \textit{may} be before the read in $P$. However, in constructing $L$, the read is pulled backwards to just after the last propagate on the reading thread. This propagate must be before the write (or else the propagate would have got the value of the write). Therefore in $L$ the read is before the write, as required.
\end{IEEEproof}

\begin{proposition}\label{ppotso}
   The order \relPPO, defined as $\relPPO \triangleq \relPO \setminus \relWR$, is a sub-order of the logical time order $L$ from Definition~\ref{logtime}.
\end{proposition}
\begin{IEEEproof}
  Note that all such events occur in the order \relPPO in $P$. Since writes are not moved, the order between write-write pairs is preserved in $L$. Since reads are moved, but are moved in program order, order between read-read pairs is preserved in $L$. 
  Since reads are only moved backwards, order between read-before-write pairs is preserved in $L$.
\end{IEEEproof}

\begin{proposition}\label{poloctso}
  The order \relPOLOC, defined as relating events from the same processor and touching the same address in program order, is a sub-order of the logical time order $L$ from Definition~\ref{logtime}.
\end{proposition}
\begin{IEEEproof}
  Note that all such events occur in the order \relPOLOC in $P$. Since the pairs in \relPOLOC are a subset of those in \relPPO (except for write-read pairs), the same proof cases as for Proposition~\ref{ppotso} apply. For the new case of write-read pairs to the same address, note that reads are never pulled backwards over a write to the same address in constructing $L$, so the order from $P$ is preserved in $L$.
\end{IEEEproof}

We can now prove Theorem~\ref{thm:tsolb_is_tso}.
\begin{IEEEproof}
We require two facts about order relations: if orders $R_1$ and $R_2$ are sub-orders of $L$, then so is their union $R_1 \cup R_2$; and if an order $R$ is a sub-order of $L$, then so is any sub-order of $R$. Using the second fact, \relFRE and \relRFE are sub-orders of \relFR and \relRF respectively, and so (using Propositions~\ref{frtso} and \ref{rftso}) are sub-orders of $L$. Using the first fact, 
$\relPROP \triangleq \relPPO \cup \relRFE \cup \relFR$, 
$\relCOM \triangleq \relCO \cup \relRF \cup \relFR$, and
$\relHB \triangleq \relPPO \cup \relRFE$ are all sub-orders of $L$. This then gives us the axioms from Alglave et al. \cite{Alglave2014a}; \textsc{SC PER LOCATION, NO THIN AIR, OBSERVATION} and \textsc{PROPAGATION} (Definitions~\ref{def:sc_per_location}, \ref{def:no_thin_air}, \ref{def:observation}, \ref{def:propagation} respectively), as the acyclicity and irreflexivities for the axioms are satisfied by any sub-orders of a strict linear order $L$.
\end{IEEEproof}

Using the framework of Alglave et al.~\cite{Alglave2011} (Appendix~\ref{apx:framework}) we have shown that all valid traces of TSO-LB are permitted by the TSO axiomatic model. 

\subsection{TSO-LB is stronger than TSO}\label{apx:counter}
We note, however, that TSO-LB is in fact subtly stricter than TSO, which we demonstrate with the litmus test in Figure~\ref{fig:nontso}. The result is allowed by TSO; specifically, considering the store-buffering model of TSO, it is possible for both loads of $a$ and $b$ to read $0$, since it is possible for the corresponding writes of $a$ and $b$ to still be held in the the local store buffers of $P1$ and $P2$ respectively, but writes of $x$ and $y$ have already propagated to the global memory. The behaviour cannot be reproduced in TSO-LB, as writes do not enter a queue and are propagated from global to local memory in ``batches.''  Indeed, the absence of queues in TSO-LB is crucial to enable finite-state model checking in our approach.

\begin{figure}
\centering
{\small{
\begin{tabular}{|p{2.3cm}|p{2.3cm}|} \hline
\multicolumn{2}{|c|}{Init: $x = y = a = b = 0$} \\ \hline
\multicolumn{1}{|c}{$P1$} &
\multicolumn{1}{|c|}{$P2$} \\ \hline
$x \leftarrow 1$ & $y \leftarrow 1$ \\
$a \leftarrow 1$ & $b \leftarrow 1$ \\
$r_1 \leftarrow y$ & $r_4 \leftarrow x$ \\
$r_2 \leftarrow y$ & $r_5 \leftarrow x$ \\
$r_3 \leftarrow b$ & $r_6 \leftarrow a$ \\\hline
\multicolumn{2}{|r|}{Observable?: $r_1 = 0 \land r_2 = 1 \land r_3 = 0$}\\
\multicolumn{2}{|r|}{$\land\ r_4 = 0 \land r_5 = 1 \land r_6 = 0$} \\\hline
\end{tabular}
}}
\caption{Litmus test result allowed by TSO, but not observable with TSO-LB. 
}
\label{fig:nontso}
\end{figure}

\end{document}